\documentclass[aps,pre,twocolumn,showpacs,superscriptaddress,amsmath,amssymb]{revtex4-1}   

\usepackage{graphicx}
\usepackage{dcolumn}
\usepackage{bm}
\usepackage{epstopdf}
\usepackage{url}
\makeatletter

\newcommand{\Rmnum}[1]{\expandafter\@slowromancap\romannumeral #1@}
\makeatother

\begin{document}


\title{Flow Distances on Open Flow Networks}


\author{Liangzhu Guo}
\affiliation{School of Systems Sciences, Beijing Normal University, Beijing, China}
\author{Xiaodan Lou}
\affiliation{School of Systems Sciences, Beijing Normal University, Beijing, China}
\author{Peiteng Shi}
\affiliation{Science and Technology on Information Systems Engineering Laboratory, National University of Defence Technology, Changsha, China}
\author{Jun Wang}
\affiliation{Science and Technology on Information Systems Engineering Laboratory, National University of Defence Technology, Changsha, China}
\author{Xiaohan Huang}
\affiliation{Science and Technology on Information Systems Engineering Laboratory, National University of Defence Technology, Changsha, China}
\author{Jiang Zhang} \email[zhangjiang@bnu.edu.cn]{} \homepage[]{http://www.swarma.org/jake}
\affiliation{School of Systems Sciences, Beijing Normal University, Beijing, China}


\date{\today}
\setcounter{secnumdepth}{2}
\begin{abstract}
Open flow network is a weighted directed graph with a source and a sink,
depicting flux distributions on networks in the steady state of an open flow
system. Energetic food webs, economic input-output networks, and
international trade networks, are open flow network models of energy flows between species, money or value flows between industrial sectors, and goods flows between countries, respectively. Flow distances (first-passage or total) between any given two nodes $i$ and $j$ are defined as the average number of transition steps of a random walker along the network from $i$ to $j$ under some conditions. They apparently deviate from the conventional random walk distance on a closed directed graph because they consider the openness of the flow network. Flow distances are explicitly expressed by underlying Markov matrix of a flow system in this paper. With this novel theoretical conception, we can visualize open flow networks, calculating centrality of each node, and clustering nodes into groups. We apply flow distances to
two kinds of empirical open flow networks, including energetic food webs and economic
input-output network. In energetic food webs example, we visualize the trophic level of each species and compare flow distances with other distance metrics on graph. In input-output network, we rank sectors according to their average distances away other sectors, and cluster sectors into different groups. Some other potential applications and mathematical properties are also discussed. To summarize, flow distance is a useful and powerful tool to study open flow systems.
\end{abstract}

\pacs{89.90.+n}

\maketitle

\section{Introduction}\label{sec.Introduction}

A large number of studies have proved that complex network is a
powerful and useful tool to model complex systems\cite{newman_2006,barabasi_scalefree_2003,strogatz_explore_2001,albert_statistical_2002}. However, due to
the limitation of the traditional graphs for describing the
complexity of the various real systems, weighted networks\cite{barrat_architecture_2004,horvath_weighted_2011}, directed
networks\cite{bangjensen_digraph_2000}, bi-partite graphs\cite{asratian_bipartite_1998}, multiplex\cite{buldyrev_catastrophic_2010,kyu_correlated_2012}, temporal networks\cite{petter_temporal_2012} as novel
extensions of the conventional graphs emerge in the past decade. Among these, open flow network is a particular kind of directed weighted network to depict
open flow system.

Most complex systems are open, they exchange energy and material
with their environment\cite{nicolis_self_1977}. Energy and material flows are delivered
to each unit of a system by the flow network\cite{west_origin_2005,banavar_size_1999}. The distribution of
these flows in the entire body of a system is
described by directed weighted edges. Two special nodes
``source'' and ``sink'' are always added in the system to represent
environment. Because the flow system considered is supposed to be in a steady
state, the flow network is always balanced which means that the total inflow
of each node equals to its total out flow except for the sink
and the source.

Energetic food web is a typical open flow network which has been
studied for several years by system ecologists. The seminal work of
H.T. Odum~\cite{odum_system_1983,odum_self-organization_1988} has depicted complicated energy flow
transactions between two species as energy circuit. A bunch of
indicators have been proposed to quantify the properties of this
open flow network\cite{fath_review_1999,finn_measures_1976,higashi_network_1993,levine_several_1980}, and numeric common properties have been
discovered\cite{ulanowicz_ecology_1997,ulanowicz_quantitative_2004,zhang_scaling_2010,zhang_allometry_2013,zhang_Common_2013}. Patten $et\ al$. proposed a systematic ``Ecological Flow Analysis'' method to investigate energetic flow networks\cite{patten_environs_1982,patten_environs_1981}.

Indeed, many basic ideas and approaches of flow analysis on energetic food webs inherit from the economic
input-output analysis method\cite{higashi_extended_1986} which is first proposed by the famous economist Leontief\cite{leontief_structure_1941,leontief_input-output_1986}. To quantify the complex economic production processes and the interaction between different economic sectors, an input-output matrix is calculated for an economic system to represent goods flows\cite{raa_economics_2005,miller_input-output_2009}. Following Leontief's seminal work, Hanon introduced basic notions such as fundamental matrix to ecology for describing the energy flows between species\cite{hannon_structure_1973}. Therefore, an input-output matrix can also be regarded as an open flow network. Money flow from the final demands compartment, circulate in different sectors of an economic system, and eventually flow to the value added compartment (or goods flow in an inverse direction). Thus, value-added compartment can be regarded as the sink, and final demands can be regarded as the source. The money flow from industry $i$ to industry $j$ is always measured by the uniform currency unit, therefore the total out flow from the source equals the total inflow to the sink, and is identical to the gross domestic output of an economy\cite{raa_economics_2005,miller_input-output_2009}. Other examples of open flow networks include clickstream networks\cite{lingfei_metabolism_2014,lingfei_clickstream_2013} and trade networks\cite{shi_trade_2014}. In summary, open flow network is a very useful tool to depict various open flow systems.

Distance on graph is a very useful concept\cite{brockmann_hidden_2013}. Both the shortest
path distance\cite{boris_path_1996}, resistance distance\cite{klein_distance_1993} and the mean first-passage distance of a random
walker\cite{noh_randomwalk_2004,blochl_vertex_2011,laszlo_random_1993,tetali_random_1991} can reflect the intrinsic properties of the graph. However,
conventional first-passage distance on a graph is
based on the basic assumption that the whole network is closed,
which means the random walker cannot escape from the network, thus
the total number of walkers on the graph is conservative.
Nevertheless, the open flow network is an open system. Random
walkers can flow into the system from the source and flow out to the
sink despite the total number of walkers staying in the network can be also conservative if the flow system is in a steady state. Therefore, the traditional method for closed system cannot be
simply extended to open flow networks. It is necessary to
extend the distance notions for open flow networks.

This paper is organized as follows. In section ~\ref{sec.Flow Distances}, the flow distance
quantities from $i$ to $j$ are defined. The explicit form of each flow
distance is expressed in sub-section \ref{sec.Calculation of flow distances}. Sub-section \ref{sec.Calculation on an example network} shows how the distance matrix is calculated on an example flow network. In sub-section \ref{sec.Food web}, we apply our method to energetic food webs, visualize each species by its trophic level, and compare different distances on the food webs. The applications of flow distances on input-output network including network visualization, sector clustering, and vertex centrality are introduced in sub-section \ref{sec.Input-output network}. Finally, we give a short summary for all the paper and the perspective of flow distances in section \ref{sec.Discussion}.

\section{Flow Distances}\label{sec.Flow Distances}
In this section, we will present the definitions and calculations of flow distances. Three flow distances, namely first-passage flow distance, total flow distance, and symmetric flow distance, are defined. They all can be expressed by the Markov matrix of the open flow network. To obtain the final expressions, some intermediate concepts including total flow and first-passage flow are needed to be introduced.

\subsection{Definitions}\label{sec.Definitions}
Consider an open flow network with $N$ common nodes and two special nodes ``source'' denoted by $0$ and ``sink'' denoted by $N+1$ are added. An $(N+2)\times (N+2)$ matrix $F$ can be used to represent flows, and each entry $f_{ij}$, where $i,j\in {0,1,2,\cdot\cdot\cdot,N+1}$, represents the flow from node $i$ to $j$. Note that the elements in the first column and the last row are all equal $0$ because there are no inflow to the source and no out flow for the sink. We also define $f_{i\cdot}=\sum_{j=0}^{N+1}f_{ij}$ as the total out flow from $i$, and $f_{\cdot j}=\sum_{i=0}^{N+1}f_{ij}$ as the total inflow to $j$. In our research, the flow network should be balanced, which means that $f_{\cdot i}=f_{i\cdot}$ for every node $i$ except ``source'' and ``sink''. Particularly, we name $f_{i,N+1}$, the flow from $i$ to the sink, as \textbf{dissipation}.

Suppose a large number of particles flow along links in the network $F$, the directed flow $f_{ij}$ from $i$ to $j$ is the total number of particles that jump from $i$ directly to $j$ along edge $i\rightarrow j$ in each time. The particles may jump from $i$ to $j$ along indirected paths, we define the \textbf{first-passage flow} from $i$ to $j$ denoted by $\phi_{ij}$ as the number of particles that reach $j$ in each time step for the first time and have been visited $i$. And the average step that these particles have jumped is defined as the \textbf{first-passage flow distance} which is denoted by $l_{ij}$.

Similarly, the \textbf{total flow} from $i$ to $j$ denoted as $\rho_{ij}$ is defined as the total number of particles that have been visited $i$ and arrive at $j$ in each time no matter if it is the first time or not. And the average step that these particles have jumped is defined as the \textbf{total flow distance} which is denoted by $t_{ij}$.

To understand these quantities better, let's consider the following imaginary experiment. Suppose all the particles passing by node $i$ are dyed red and this color would be washed out once the red particles arrive at node $j$ for the first time. Then the first-passage flow from node $i$ to $j$ is the number of red particles passing by node $j$ in each time. The first-passage flow distance is the average step that these particles have made. Similarly, if the particles passing by $i$ are dyed red but this color would never be washed out, then the number of red particles that pass by $j$ in each time is the total flow, and the average step that these particles have made is the total flow distance. 

In this paper, all the matrices are denoted by capital letters, and the their corresponding elements are denoted by lower case of the name of matrices. For example, $F$ denote the flow matrix, and $f_{ij}$ is the element of the $i$th row and $j$th column

\subsection{Calculation of total flow and first-passage flow}\label{sec.Calculation of total flow and first-passage flow}
Because the open flow system is in a steady state, and the flow network is balanced, we can define a Markov matrix $M$ as follows,
\begin{equation}
\label{eqn.Markov}
m_{ij}=\frac{f_{ij}}{\sum_{j=1}^{N+1}f_{ij}}
\end{equation}
and $m_{ij}$ represents the probability of particles jumping from state $i$ to $j$. Note that $\sum_{j=1}^{N+1}m_{ij}=1$ for any $i$ except $N+1$ because the elements in the last ($(N+1)$th) row are all zeros, this is a key difference between open flow network and closed flow network.

According to reference \cite{higashi_network_1993}, no matter if circulations exist in network, the total flow from $i$ to $j$ can be calculated as:
\begin{equation}
\label{eqn.totalflow}
\rho_{ij}=\phi_{0i}u_{ij},
\end{equation}
where
\begin{equation}
\label{eqn.fundamentalmatrix}
U=I+M+M^2+\cdot\cdot\cdot=(I-M)^{-1}
\end{equation}
is called fundamental matrix, it is also the inverse of $M$'s laplacian. And $\phi_{0i}$ is the first-passage flow from the source to $i$ which will be calculated in the following paragraphs. $I$ is the identity matrix with size $(N+1) \cdot (N+1)$. We will provide an informal proof for Eq~(\ref{eqn.totalflow}).

Equation (\ref{eqn.totalflow}) calculates the total flows along all possible paths from $i$ to $j$. When $i\neq j$, the number of particles that jump from $i$ to $j$ along all possible paths with $k$ steps is $\phi_{0i}(M^k)_{ij}$. Note that particles may flow back to $i$ for several times, $\phi_{0i}$ instead of $f_{\cdot i}$ is adopted because $f_{\cdot i}$ contains the flows back to $i$. If the particle passing by $i$ is dyed, then $\phi_{0i}$ is the number of particles without color marker and will be dyed in each time. Taking summation of $\phi_{0i}(M^k)_{ij}$ from $k=1$ to $\infty$, we can obtain the total flow from $i$ to $j$ along all possible ways. According to the series expansion $MU=M(I-M)^{-1}=M+M^2+\cdot\cdot\cdot$ and the identity $(MU)_{ij}=u_{ij}$ when $i\neq j$, Eq.~(\ref{eqn.totalflow}) is obtained.

When $i=j$, according to $\rho$'s definition, $\rho_{ii}$ should contain the first-passage flow from the source to $i$, therefore, $\rho_{ii}=\phi_{0i}((MU)_{ii}+1)=\phi_{0i}((MU)_{ii}+I_{ii})=\phi_{0i}u_{ii}$, then Eq.~(\ref{eqn.totalflow}) holds.

Because the total flow from $i$ to $j$ can be divided into two different categories, one is the first-passage flow which contains the particles that arrive at $j$ for the first time, the other is the circulation flow which contains the particles that arrive at $j$ more than once. All the flows are conditioned on starting from $i$. We know that the circulation flow is the summation of flows from $j$ to $j$ along all possible paths, it is calculated as
\begin{equation}
\psi_{ij}=\phi_{ij}(\sum_{k=1}^{\infty}M^k)_{jj}=\phi_{ij}(MU)_{jj},
\end{equation}
where $\psi_{ij}$ represents the circulation flow starting from $i$.

Therefore, the total flow from $i$ to $j$ can be expressed as\cite{higashi_network_1993}
\begin{equation}
\rho_{ij}=\phi_{ij}+\psi_{ij}=\phi_{ij}u_{jj}.
\end{equation}
Thus, we obtain the expression for the first-passage flow from $i$ to $j$:
\begin{equation}
\label{eqn.firstpassageflow}
\phi_{ij}=\frac{\rho_{ij}}{u_{jj}}.
\end{equation}

Based on the equations of Eq.~(\ref{eqn.totalflow}) and Eq.~(\ref{eqn.firstpassageflow}), and note that $\phi_{00}=f_{0\cdot}$ according to the definition, where $f_{0\cdot}$ denotes the total flow from ``source'' to the whole system, we have
\begin{equation}
\label{eqn.finaltotalflow}
\rho_{ij}=\phi_{0i}u_{ij}=\frac{\rho_{0i}}{u_{ii}}u_{ij}=f_{0\cdot}\frac{u_{0i}}{u_{ii}}u_{ij}.
\end{equation}

And the explicit expression for the first-passage flow is
\begin{equation}
\label{eqn.finalfirstflow}
\phi_{ij}=\frac{\rho_{ij}}{u_{jj}}=\phi_{0i}u_{ij}\frac{1}{u_{jj}}=f_{0\cdot}\frac{u_{0i}u_{ij}}{u_{ii}u_{jj}}.
\end{equation}

\subsection{Calculation of flow distances}\label{sec.Calculation of flow distances}
We can deduce the explicit expression of various flow distances once the total flow and first-passage flow expressions are given. First, according to the definition of the total flow from $i$ to $j$ along all possible paths, we have
\begin{equation}
\label{eqn.defoftij}
t_{ij}=\sum_{k=1}^{\infty}kp_{ij}^{k},
\end{equation}
where $p_{ij}^{k}$ denotes the probability that particles transfer from $i$ to $j$ after $k$ steps. One may think $p_{ij}^k=(M^k)_{ij}$, however, it is not true because $p_{ij}^k$ is normalized for all paths with all possible lengths $k$, i.e., $\sum_{k=1}^{\infty}p_{ij}^k=1$. However, $(M^k)_{ij}$ is normalized for all $j$s, i.e., $\sum_{j=0}^{N+1}(M^k)_{ij}=1$. We know that the flow from $i$ to $j$ after $k$ steps is $\phi_{0i}(M^k)_{ij}$ and the total flow along all possible paths is $\rho_{ij}$, therefore,
\begin{equation}
\label{eqn.pijk}
p_{ij}^{k}=\frac{\phi_{0i}(M^{k})_{ij}}{\rho_{ij}}.
\end{equation}
Thus bring this equation to Eq.~(\ref{eqn.defoftij}), we have,

\begin{align}
\label{eqn.tij}
t_{ij}&=\sum_{k=1}^{\infty}k\frac{\phi_{0i}(M^{k})_{ij}}{\rho_{ij}}=\frac{\phi_{0i}(\sum_{k=1}^{\infty}kM^{k})_{ij}}{\rho_{ij}}\notag\\
&=\frac{\phi_{0i}(MU^{2})_{ij}}{\rho_{ij}}=\frac{\phi_{0i}(MU^{2})_{ij}}{\phi_{0i}u_{ij}} \notag\\
&=\frac{(MU^{2})_{ij}}{u_{ij}}.
\end{align}
In which, we have used the following series expansion:
\begin{align}
\label{eqn.series}
MU^2=M\left(\frac{1}{I-M}\right)^2=\sum_{k=1}^{\infty}kM^k.
\end{align}

Similarly, we can obtain the expression for first-passage flow distance. First, according to the definition of the first-passage distance from $i$ to $j$, we have
\begin{equation}
\label{eqn.deflij}
l_{ij}=\sum_{k=1}^{\infty}kq_{ij}^{k},
\end{equation}
where $q_{ij}^{k}$ denotes the probability that particles started from $i$ to $j$ after $k$ steps in the first time. One cannot use $p_{ij}$ (Eq.~(\ref{eqn.pijk})) because it contains the circulation flow from $j$ to $j$. Let us assume that all the particles arriving at $j$ will be removed from the system, that is to say, we assume that $j$ is another sink, then all the calculations for the total flow distance is correct. To make this point clear, we define a new matrix $M_{-j}$ as:
\begin{equation}
(M_{-j})_{rs}=
\left\{
\begin{array}{ll}
m_{rs},& r\neq j\\
0,& r=j.
\end{array}
\right.
\end{equation}
And the correct expression for the probability $q_{ij}^k$ is
\begin{equation}
q_{ij}^{k}=\frac{\phi_{0i}(M_{-j}^{k})_{ij}}{\phi_{ij}}.
\end{equation}
Insert it into Eq.~(\ref{eqn.deflij}), we have

\begin{equation}
l_{ij}=\frac{u_{jj}(M_{-j}U_{-j}^{2})_{ij}}{u_{ij}}.
\end{equation}

According to the Theorem 1 proved in the Supplementary Material, when $u_{ij}\neq 0$ ($i$ connects to $j$), this formula can be reduced to
\begin{equation}
l_{ij}=\frac{(MU^{2})_{ij}}{u_{ij}}-\frac{(MU^{2})_{jj}}{u_{jj}}=t_{ij}-t_{jj}.
\label{eqn.firstpassagedistance}
\end{equation}
Therefore, the difference between $t_{ij}$ and $l_{ij}$ is just the total flow distance from $j$ to $j$. The matrix $\left(t_{jj}\right)_{N+1,N+1}$ have identical rows. Therefore, we can use a vector $t_j$ to abbreviate the matrix $t_{jj}$, it quantifies the ability of self circulation of each node in the system.

All the flow distances introduced above are asymmetric, however, some real tasks such as nodes clustering, computation of node centrality require symmetric metrics. Commute distance\cite{tetali_random_1991,laszlo_random_1993} is a classical and famous symmetric distance measure defined by random walk, which is calculated by $l_{ij}+l_{ji}$. However, this definition cannot work when one of $l_{ij}$ or $l_{ji}$ is infinity meaning that $i$ cannot access $j$ or vice versa. Therefore, we define a new symmetric flow distance to avoid this problem:
\begin{equation}
\label{eqn.symmetricdistance}
c_{ij}=2\frac{1}{\frac{1}{l_{ij}}+\frac{1}{l_{ji}}}=\frac{2l_{ij}l_{ji}}{l_{ij}+l_{ji}}.
\end{equation}
We call $c_{ij}$ symmetric flow distance, it is a mixing of $l_{ij}$ and $l_{ji}$. Suppose $l_{ij}=\infty$, then $c_{ij}=2l_{ji}$ which is well-defined. When $l_{ij}=l_{ji}$, $c_{ij}=l_{ij}=l_{ji}$. Therefore, $c_{ij}$ is a reasonable symmetric distance.

\subsection{Calculation on an example network}\label{sec.Calculation on an example network}

\begin{figure}[h!]
\includegraphics[scale=0.4]{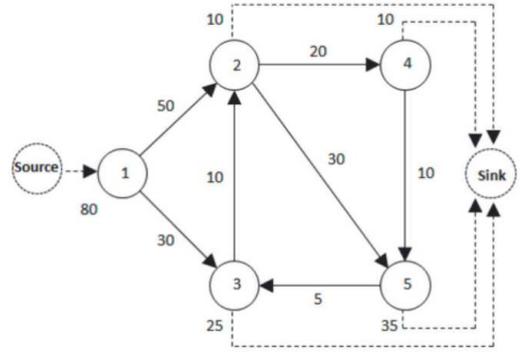}
\caption{An example open flow network}
\label{fig.examplenetwork}
\end{figure}

Before applying our method to real open flow networks, we would like to present the computations of flow distances on a small example network and compare with other distances on graph. The example network is shown in Figure \ref{fig.examplenetwork}. There are $7$ nodes including the source and the sink. All the flows are denoted on the edges. We present the first-passage flow distances matrix $L$ in the following equation

\begin{equation}
\label{eq.exampleL}
\left[
\begin{array}{ccccccc}
0&1&2.15&2.27&3.20&3.40&3.94\\
\infty&0&1.15&1.27&2.20&2.40&2.94\\
\infty&\infty&0&2.25&1.05&1.25&2.13\\
\infty&\infty&1&0&2.05&2.25&1.61\\
\infty&\infty&3&2&0&1&1.60\\
\infty&\infty&2&1&3.05&0&1.20\\
\infty&\infty&\infty&\infty&\infty&\infty&0
\end{array}
\right]
\end{equation}
and the total flow distances matrix $T$\\
\begin{equation}
\label{eqn.exampleT}
\left[
\begin{array}{ccccccc}
0&1&2.23&2.35&3.23&3.48&3.94\\
\infty&0&1.23&1.35&2.23&2.48&2.94\\
\infty&\infty&0.08&2.33&1.08&1.33&2.13\\
\infty&\infty&1.08&0.08&2.08&2.33&1.61\\
\infty&\infty&3.08&2.08&0.03&1.08&1.60\\
\infty&\infty&2.08&1.08&3.08&0.08&1.20\\
\infty&\infty&\infty&\infty&\infty&\infty&0
\end{array}
\right]
\end{equation}

Note that there are many $\infty$ entries in both $L$ and $T$ because the corresponding node pairs have no connected path. Another interesting phenomenon is all the elements in $T$ are larger than the corresponding entries in $L$. And the difference $(T-L)$ is:
\begin{equation}
\left[
    \begin{array}{ccccccc}
        0&0&0.08&0.08&0.03&0.08&0\\
        &0&0.08&0.08&0.03&0.08&0\\
        &&0.08&0.08&0.03&0.08&0\\
        &&0.08&0.08&0.03&0.08&0\\
        &&0.08&0.08&0.03&0.08&0\\
        &&0.08&0.08&0.03&0.08&0\\
        &&&&&&0
        \end{array}
\right]
\end{equation}

The empty entries have no numeric value because $\infty-\infty$ is indefinite. All the elements in the same column are identical which are the average flow distances from $i$ to $i$ for $i=2,3,4,5$. And because $2,3,5$ are in the same cycle $2\rightarrow5\rightarrow3$ and $2\rightarrow4\rightarrow5\rightarrow3$, they have the same values of $t_{jj}$.

Next, we compare our first-passage flow distance $l_{ij}$ with shortest path distance and first-passage distance based on random walks\cite{blochl_vertex_2011} on the closed version of the same network. In the latter comparison, ``source'' and ``sink'' are excluded so that the network is closed. For the random walkers in the closed network, the transition probability between $i$ and $j$ is the fraction between $f_{ij}$ and ($f_{i\cdot}-f_{i,N+1}$), the total out flows from $i$ excluding the dissipation from $i$ . For example, the transition probability from $2$ to $4$ is $20/(20+30)=0.4$ but not $20/(20+30+10)=0.33$. The results are shown in Table \ref{tab.example}.

\begin{table}[h]
\caption{Comparisons among three kinds of distances on selected node pairs}
\begin{tabular}{lcccc}
\hline
\hline
 &1$\rightarrow$3&2$\rightarrow$3&1$\rightarrow$4&2$\rightarrow$4\\
 \hline
Shortest path&1&2&2&1\\
Closed FPD&2.5&2.4&6.875&5.5\\
Open FPD&1.274&2.25&2.2&1.055\\
\hline
\hline
\end{tabular}
\label{tab.example}
\end{table}

As we expected, shortest path lengths are much shorter than the other two distances because they only consider the shortest paths, as a result, this distance always under-estimates the distances of random particles in flows. Comparing two first-passage distances is more interesting. Closed First Passage Distance (Closed FPD) is always larger than Open First Passage Distance (Open FPD) because dissipations are not considered in Closed FPD. For example, the Closed FPD from $1$ to $4$ is larger than the Open FPD in almost $3$ times because the longer circulated path $2\rightarrow 5\rightarrow 3\rightarrow 2$ has much higher probability when the dissipation from node $2$ is neglected ($0.6$ rather than $0.5$).

\section{Empirical Studies}\label{sec.Empirical Studies}

In this section, we will apply our flow distances on two kinds of networks: $18$ energetic food webs (energy flow information between species is included) and the input-output network of U.S. The raw data of food webs is from the following open data source\cite{foodwebs_address}. And the input-output network data is from\cite{io_address}.
\subsection{Food web}\label{sec.Food web}
\begin{figure*}
\includegraphics[scale=1.7]{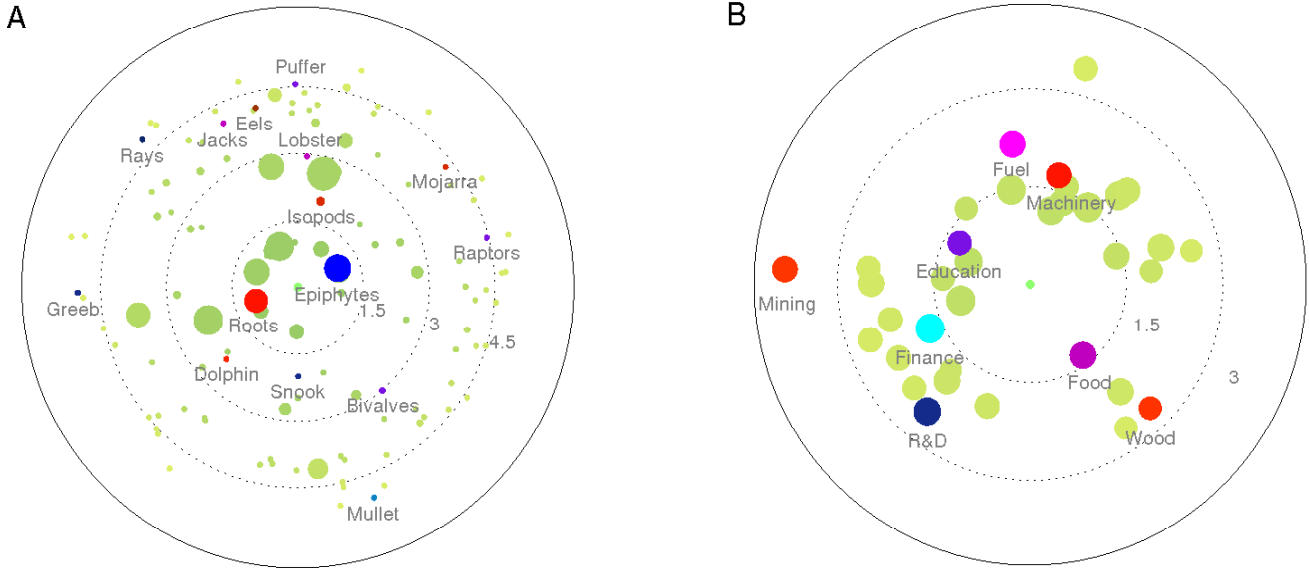}
\caption{Trophic levels of species in Baydry community (a) and industrial sectors of U.S. input-output network in 2000 (b). The polar radii, i.e., the distances between every node and the center are proportional to nodes' trophic levels, the polar angles are randomly assigned, and the sizes of nodes are proportional to the logarithmic volumes of the total throughflow for each node ($f_{i\cdot}$). The colors are assigned randomly.}
\label{fig.trophiclevels}
\end{figure*}

Trophic level is an important concept in food webs, it characterizes one species' distance from the source (sun light) along the food chain. However, when the food web is an entangled network, calculating the shortest path from the source always under-estimates the trophic level of a given species because non-shortest paths may have much longer distances from the source. Therefore, we quantify trophic levels of different species by the concept of first-passage distance from the source\cite{levine_several_1980}, that is $l_{0i}$ for any species $i$. This distance is reasonable because it contains the information of weights and all possible energetic path ways from the source. Figure~\ref{fig.trophiclevels}.a visualizes the trophic levels of $125$ biological species in Baydry food web. Producer species locate at the area close to the source, and higher level consumers locate in the peripheries. \\
Next, we calculate several distances in the level of the entire network. The first distance is the first-passage distance from the source to the sink ($l_{0,N+1}$). This distance quantifies the average number of steps of a random particle in its all life span. The second distance is the mean value of the elements in matrix $L$ except for the infinite elements. We calculate these distances for all the collected energetic food webs, and to observe how the distances change with network size.\\
\begin{figure}[h]
\includegraphics[width=0.53\textwidth,angle=0]{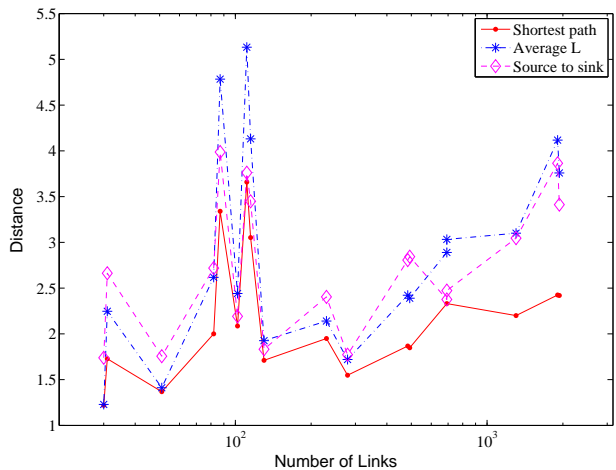}
\caption{Three kinds of distances for all collected energetic food webs. All food webs are sorted according to their number of edges in an increasing order}
\label{fig.distancesfoodweb}
\end{figure}
Figure~\ref{fig.distancesfoodweb} shows various distances change with number of edges of networks. We find that the average value of $l_{ij}$ has similar trend with the average path length $l_{0,N+1}$ from the source to the sink. Shortest path length is always shorter than the average $l$ and $l_{0,N+1}$ because it does not consider the average behaviour of random walkers. There is a slightly trend that the network lengths increase with network size.
\subsection{Input-output network}\label{sec.Input-output network}
Input-output network is another kind of flow network. Each industrial sector corresponds to a vertex, and an input from one sector to another can be considered as a flow. However, there are two kinds of views to represent an input-output network as a flow network. If we consider material flow, then the input from sector $i$ to $j$ should be understood as a flow from $i$ to $j$. However the flow may be from $j$ to $i$ if money flow is considered. We adopt the viewpoint of money flow in this paper because the flow of money in different sectors resembles random walkers in open flow networks. In this way, the final demand sector is the source of money flows, and the value added sector is the sink. We choose the input-output data from United States in 2000 as an example to calculate various flow distances.

First, it is curious to calculate the economic ``trophic levels'' ($l_{0,i}$) of different sectors (see Figure~\ref{fig.trophiclevels}.b). The sectors with shorter distances from the center are closer to the source, therefore they are more easily to be affected by the final demand. Any fluctuations of demands or price can be transferred to the sectors with lower ``trophic levels''.

\begin{figure}[h]
\includegraphics[scale=0.6]{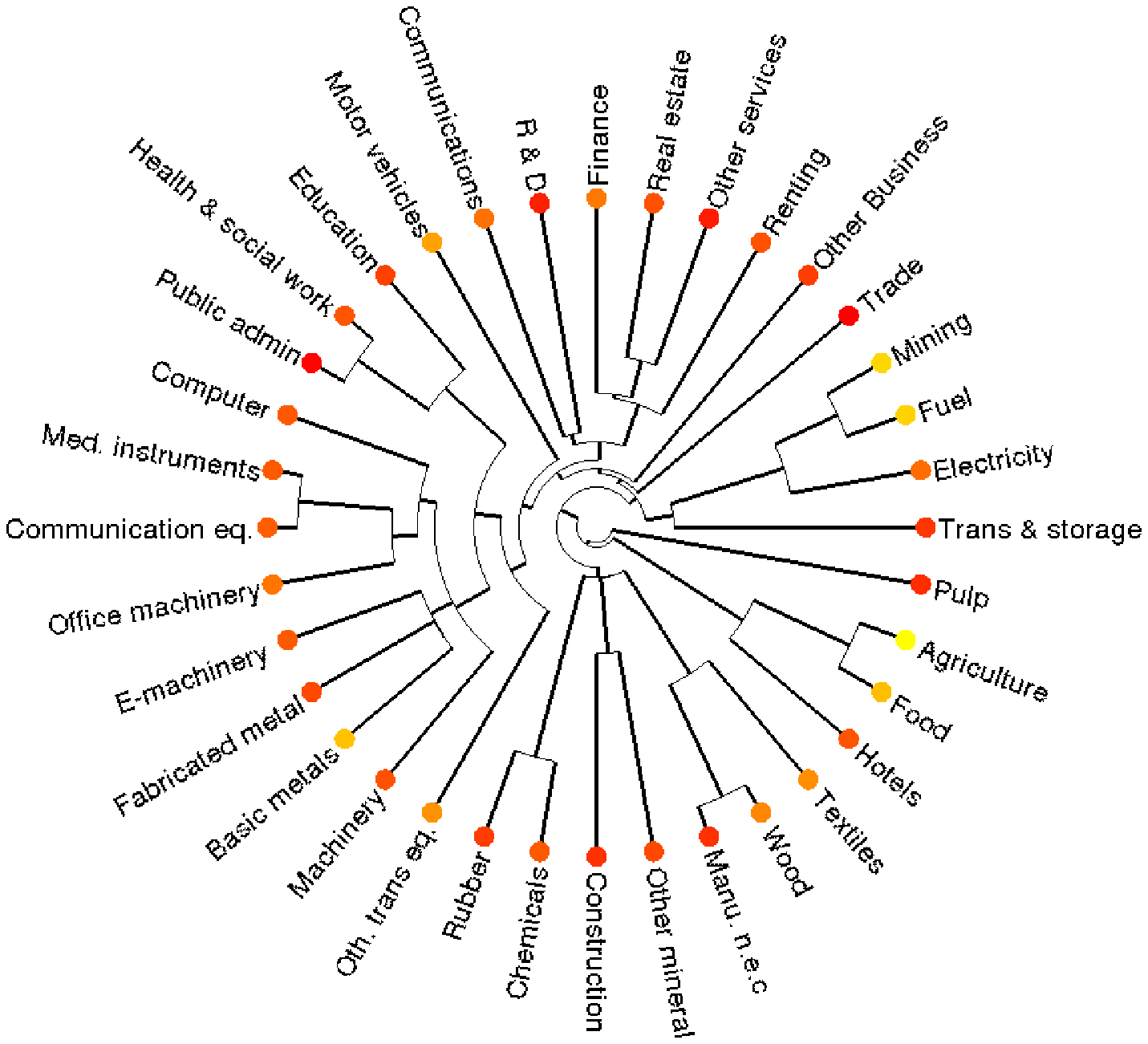}
\caption{Hierarchical clustering of different industrial sections in U.S. Colors represent node centrality. All the sector names are abbreviated, and full names can be referred to the Supplementary Material.}
\label{fig.cluster}
\end{figure}

Second, we can use flow distances to calculate similarity between different sectors. Because it is much easier to deal with symmetric similarity, we use the distances $c_{ij}$ instead of $l_{ij}$ here. With this symmetric measure, we can cluster sectors by using the standard hierarchical clustering techniques\cite{Johnson_hierarchical_1967}. The result is visualized by Figure~\ref{fig.cluster}.
In this figure, similar or related sectors are gathered closely, like $Public\ admin$ and $Health\ \&\ social work$, $Ming$ and $Fuel$. We also find that $Real\ estate$ sector is close to $Finance$ sector, which means real estate has tight relation with finance in U.S. The clustering results have good agreement with our common sense of industrial sectors.

Furthermore, the symmetric measure $c_{ij}$ can be used to measure the centrality of each node because if $i$'s average $c_{ij}$ for different $j$ is shorter then $i$ must have tight connections with all other nodes. Formally, we define the centrality of node $i$ as
\begin{equation}
\bar{c_{i}}=\frac{\sum_{j=1}^{N}c_{ij}}{N}
\end{equation}
Thus, the shorter is $i$'s $\bar{c_{ij}}$, the more central position it has in the whole economic system.
We color different nodes in Fig.~\ref{fig.cluster} by $\bar{c_i}$. The color depth increases as $c_i$ decreases. We find that $Trade$ and $Public\ admin.$ sectors are more central than other sectors in U.S., and $Agriculture$ and $Ming$ sectors are less important than the average.

Finally, we calculate the vector $t_{i}$ for all $i$. It is defined as the average steps of a random walker who starts from $i$ and finally returns to $i$ again. This measure indicates the re-cycle capability of a sector in the sense of money flow. Therefore, less $t_{i}$ implies larger capability of self-maintenance of this sector. In Table~\ref{tab.tii}, we show the top $5$ and bottom $5$ sectors in the decreasing order of $t_{i}$ in the United States.

\begin{table}[h]
\caption{List of sectors sorted by $t_{i}$}
\begin{tabular}{cl}
\hline
\hline
 Rank&Sectors in USA\\
 \hline
1&Motor vehicles, trailers and semi-trailers\\
2&Finance and insurance\\
3&Basic metals\\
4&Chemicals and chemical products\\
5&Agriculture, hunting, forestry and fishing\\
$\vdots$&\\
32&Electricity, gas and water supply\\
33&Hotels and restaurants\\
34&Construction\\
35&Education\\
36&Health and social work\\
\hline
\hline
\end{tabular}
\label{tab.tii}
\end{table}

The top five sectors are more likely connected to other sectors in the economy. Through analysing the flux matrix $F$, we find that they have less fractions of flows from the source or to the sink. On the contrary, the last five sectors are all major in providing services or products for final demand.

\section{Discussion}\label{sec.Discussion}
In this paper, we introduce flow distances in various open flow networks. These distances characterize interactions between different nodes, and all the distances can be expressed explicitly by the Markov matrix. We give several examples on potential applications of flow distances on energetic food webs and input-output network. Trophic level as an important conception introduced in food web ecology should be applied to other open flow networks. Usually, the nodes with lower trophic levels are more probable to be influenced by the source easily. Second, we can use flow distances to cluster nodes because the symmetric distance $c_{ij}$ can be regarded as a kind of similarity measure. We also use $c_{ij}$ to compare node centrality between different nodes. Vector $t_{i}$ can be used as an indicator to compare the in-dependency of different node. Because all the flow distances reflect the nature of random walk in an open flow network, they combine the topology and flow dynamics on the network together. Therefore, these distances must have very wide application background.

Certainly, the applications of flow distances should not be limited by the examples listed in this paper. First, open flow networks are combinations of network structure and random walk dynamics, thus visualizing these networks needs particular method. Besides placing different nodes on a space by their ``trophical levels'' directly, we can embed the flow network into a Euclidean space according to distance $c_{ij}$ such that the Euclidean distance of any given pair $i$ and $j$ is as close as their $c_{ij}$. This embedding problem can be solved by optimizing the places of each node in the Euclidean space. And the patterns of the nodes distributed in the space may help us to understand the flow network structure in an intuitive way. However, how to visualize the open flow networks to reflect the characteristics of the directionality and weights of edges is another important issue deserving for further studies. Second, open flow networks always resemble tree structures that are hierarchical and possessing multi-level structures. How to partition a flow network into several smaller sub-structures, and how to coarse-grain these structures is also an interesting problem. It is reasonable to develop a novel method based on flow distances discussed in this paper to partition and coarse-grain. Third, the flow distances metrics and network embedding can help us to understand some underlying dynamical processes on the network in a geometric way\cite{brockmann_hidden_2013}.

Flow distances can obviously applied to other open flow networks, and may facilitate us to compare them. Trade flow network, traffic flow network, attention flow networks are all very important examples. Application of flow distances on these networks may reveal important common patterns.

The current flow distances metrics also have shortcomings. The computational complexity will increase fast as the size of the network because the matrices $U$ and $L$ are non-sparse when the network is large. Therefore, the approximate algorithm of flow distances is very necessary and urgent. Additionally, all the flow distances metrics are average values of various paths of particles, the variances of these paths cannot be reflected on these metrics. New indicators are needed to represent the fluctuations of different paths. All these problems deserve further studies.
\bibliography{ecology}

\end{document}



\title{Supplementary Material for Flow Distances on Open Flow Networks}


\author{Liangzhu Guo}
\affiliation{School of Systems Sciences, Beijing Normal University, Beijing, China}
\author{Xiaodan Lou}
\affiliation{School of Systems Sciences, Beijing Normal University, Beijing, China}
\author{Peiteng Shi}
\affiliation{Science and Technology on Information Systems Engineering Laboratory, National University of Defence Technology, Changsha, China}
\author{Jun Wang}
\affiliation{Science and Technology on Information Systems Engineering Laboratory, National University of Defence Technology, Changsha, China}
\author{Xiaohan Huang}
\affiliation{Science and Technology on Information Systems Engineering Laboratory, National University of Defence Technology, Changsha, China}
\author{Jiang Zhang} \email[zhangjiang@bnu.edu.cn]{} \homepage[]{http://www.swarma.org/jake}
\affiliation{School of Systems Sciences, Beijing Normal University, Beijing, China}


\date{\today}
\setcounter{secnumdepth}{2}

\pacs{}

\maketitle
\section{Proof of a theorem}\label{appendix_delete}
In this appendix, we will prove Eq.~(17). But before that, several lemmas are needed to be proved at first.
\\

\textbf{Lemma 1}: The following equation is true:
\begin{equation}
I=(I-M)U.
\end{equation}
\textbf{Proof}: It is obvious according to the definition of $U=(I-M)^{-1}$.
\\

\textbf{Lemma 2}: The following equation is true:
\begin{equation}
(I-M)U=(I-M_{-d})U_{-d}=I.
\end{equation}
Where, $M_{-d}$ is the matrix when the $d$th row of matrix $M$ is set to zero. Thus,
\begin{equation}
M=M_{-d}+\Delta M,
\end{equation}
where
\begin{equation}
(\Delta M)_{ij}=
\left\{
\begin{array}{ll}
m_{ij},& i=d\\
0,& i\neq d
\end{array}
\right.
\end{equation}
Correspondingly, $U_{-d}$ is
\begin{equation}
U_{-d}=I+M_{-d}+M_{-d}^{2}+\cdots=(I-M_{-d})^{-1}.
\end{equation}
\textbf{Proof}: This is also obvious according to Lemma 1.
\\

\textbf{Lemma 3}: The following equation holds for any $d,i,j$ belongs to $[1,N]$:
\begin{align}
(U_{-d})_{ij}&=u_{ij}-(U_{-d})_{id}u_{dj}\\
&=u_{ij}-\frac{u_{id}}{u_{dd}}u_{dj}-\frac{u_{id}}{u_{dd}}\delta_{dj}.
\end{align}
Where,
\begin{equation}
\delta_{ij}=
\left\{
\begin{array}{ll}
1,& i=j\\
0,& i\neq j
\end{array}
\right.
\end{equation}
\textbf{Proof}: According to Lemma 2, we have\\
\begin{equation}
U-MU=U-(M_{-d}+\Delta M)U=U_{-d}-M_{-d}U_{-d}
\end{equation}
\begin{equation}
\begin{aligned}
\Rightarrow U-U_{-d}&=(M_{-d}+\Delta M)U-M_{-d}U_{-d}\notag\\
&=M_{-d}(U-U_{-d})+\Delta MU\notag\\
\end{aligned}
\end{equation}
\begin{equation}
\Rightarrow (I-M_{-d})(U-U_{-d})=\Delta MU
\end{equation}
\begin{equation}
\because\ U_{-d}(I-M_{-d})=I
\end{equation}
\begin{equation}
\therefore\ U-U_{-d}=U_{-d}\cdot\Delta M\cdot U
\end{equation}
According to the definition of $\Delta M$, and according to the fact that
\begin{equation}
\sum_{k}m_{ik}u_{kj}=u_{ij}-\delta_{ij},
\end{equation}
we can expand $U_{-d}\cdot \Delta M\cdot U$ as
\begin{align}
U_{-d}\cdot (\Delta M\cdot U)_{ij}&=(U_{-d})_{id}\cdot \sum_{k}m_{dk}u_{kj}\notag\\
&=(U_{-d})_{id}(u_{dj}-\delta_{dj}).
\end{align}
So, we get
\begin{equation}
\begin{aligned}
\label{eqn.udij}
u_{ij}-(U_{-d})_{ij}=(U_{-d})_{id}(u_{dj}-\delta_{dj})
\end{aligned}
\end{equation}
In the above equation, if we let $j=d$, then
\begin{equation}
u_{id}-(U_{-d})_{id}=(U_{-d})_{id}(u_{dd}-1)
\end{equation}
Thus,
\begin{equation}
(U_{-d})_{id}=\frac{u_{id}}{u_{dd}}.
\end{equation}
Insert it into Equation~(\ref{eqn.udij}), we have
\begin{equation}
u_{ij}-(U_{-d})_{ij}=\frac{u_{id}}{u_{dd}}(u_{dj}-\delta_{dj}).
\end{equation}
At last, rearrange this equation, we obtain
\begin{equation}
(U_{-d})_{ij}=u_{ij}-\frac{u_{id}}{u_{dd}}u_{dj}-\frac{u_{id}}{u_{dd}}\delta_{dj}.
\end{equation}

\textbf{Lemma 4}: Based on these lemmas, we can get such equation:
\begin{equation}
(U_{-j}^{2})_{ij}=\frac{(U^{2})_{ij}}{u_{jj}}-\frac{u_{ij}}{u_{jj}^{2}}(U^{2})_{jj}+\frac{u_{ij}}{u_{jj}}.
\end{equation}
\textbf{Proof}: Expand $U^{2}$ into elements, and substitute Lemma 3 into it, then
\begin{align}
(U_{-j}^{2})&=\sum_{k}(U_{-j})_{ik}(U_{-j})_{kj} \notag \\
&=\sum_{k}(u_{ik}-\frac{u_{ij}}{u_{jj}}u_{jk}+\frac{u_{ij}}{u_{jj}}\delta_{jk})\frac{u_{kj}}{u_{jj}}\notag\\
&=\sum_{k}\frac{u_{ik}u_{kj}}{u_{jj}}-\sum_{k}\frac{u_{ij}}{u_{jj}^{2}}u_{jk}u_{kj}+\frac{u_{ij}}{u_{jj}}\notag\\
&=\frac{(U^{2})_{ij}}{u_{jj}}-\frac{u_{ij}}{u_{jj}^{2}}(U^{2})_{jj}+\frac{u_{ij}}{u_{jj}}.
\end{align}\\

\textbf{Theorem 1}: Equation (17) in the main text or the following equation
\begin{equation}
\frac{1}{u_{ij}}[(MU^{2})_{ij}-u_{jj}(M_{-j}U_{-j}^{2})_{ij}]=\frac{(MU^{2})_{jj}}{u_{jj}}
\end{equation}
holds when $u_{ij}\neq 0$.

\textbf{Proof}:
Substitute $M\cdot U^2=U^{2}-U$ and $M_{-j}\cdot U_{-j}^{2}=U_{-j}^{2}-U_{-j}$ into Lemma 1 and Lemma 3, it can be proved.\\
\begin{align}
\frac{1}{u_{ij}}[&(MU^{2})_{ij}-u_{jj}(M_{-j}U_{-j}^{2})_{ij}]\notag\\
&=\frac{1}{u_{ij}}\bigl[(U^{2}-U)_{ij}-u_{ij}(U_{-j}^{2}-U_{-j})_{ij}\bigr] \notag \\
&=\frac{1}{u_{ij}}\bigg\{(U^{2})_{ij}-u_{ij}-u_{jj}\biggl[\frac{(U^{2})_{ij}}{u_{jj}}-\frac{u_{ij}}{u_{jj}^{2}}(U^{2})_{jj}\biggr])\bigg\} \notag\\
&=\frac{1}{u_{ij}}\bigl[(U^{2})_{ij}-u_{ij}-(U^{2})_{ij}+\frac{u_{ij}}{u_{jj}}(U^{2})_{jj}\bigr]\notag\\
&=\frac{(U^{2})_{jj}-u_{jj}}{u_{jj}}\notag\\
&=\frac{(MU^{2})_{jj}}{u_{jj}}
\end{align}

\section{Name list for sectors of input-output network}\label{appendix_sectorname}

Sector names in Figure 4 are abbreviated. The full names corresponded are depict in Table \ref{tab.sectornames}.

\begin{table*}[h]
\caption{Sector full names}
\begin{tabular}{ll}
\hline
\hline
 Abbreviations&Full names\\
 \hline
Agriculture&Agriculture, hunting, forestry and fishing\\
Mining&Mining and quarrying\\
Food&Food products, beverages and tobacco\\
Textiles&Textiles, textile products, leather and footwear\\
Wood&Wood and products of wood and cork\\
Pulp&Pulp, paper, paper products, printing and publishing\\
Fuel&Coke, refined petroleum products and nuclear fuel\\
Chemicals&Chemicals and chemical products\\
Rubber&Rubber and plastics products\\
Other mineral&Other non-metallic mineral products\\
Basic metals&Basic metals\\
Fabricated metal&Fabricated metal products except machinery and equipment\\
Machinery&Machinery and equipment n.e.c \\
Office machinery&Office, accounting and computing machinery\\
E-machinety&Electrical machinery and apparatus n.e.c\\
Communication eq.&Radio, television and communication equipment\\
Med. instruments&Medical, precision and optical instruments\\
Motor vehicles& Motor vehicles, trailers and semi-trailers\\
Oth. trans eq.&Other transport equipment\\
Manu. n.e.c&Manufacturing n.e.c; recycling\\
Electricity&Electricity, gas and water supply\\
Construction&Construction\\
Trade&Wholesale and retail trade; repairs\\
Hotels&Hotels and restaurants\\
Trans \& storage&Transport and storage\\
Communications&Post and telecommunications\\
Finance&Finance and insurance\\
Real estate&Real estate activities\\
Renting&Renting of machinery and equipment\\
Computer&Computer and related activities\\
R\&D&Research and development\\
Other Business&Other Business Activities\\
Public admin&Public admin. and defence; compulsory social security\\
Education&Education\\
Health \& social work&Health \& social work\\
Other services&Other community, social and personal services\\
Private households&Private households with employed persons\\
\hline
\hline
\end{tabular}
\label{tab.sectornames}
\end{table*}